\documentclass[11pt,a4paper]{article}

\usepackage[T1]{fontenc}
\usepackage[utf8]{inputenc}
\usepackage{lmodern}
\usepackage{microtype}
\usepackage{xurl}
\usepackage{geometry}
\usepackage{setspace}
\usepackage{csquotes}
\usepackage{hanging}
\usepackage{amsmath,amssymb}
\usepackage{graphicx}
\usepackage{caption}
\usepackage{subcaption}
\usepackage{booktabs}
\usepackage{float}
\usepackage{enumitem}
\usepackage[colorlinks=true,
            linkcolor=black,
            citecolor=black,
            urlcolor=cyan]{hyperref}
\usepackage[numbers,sort&compress,round]{natbib}
\usepackage{array}
\usepackage{float}
\usepackage{wrapfig}

\begin{document}
\thispagestyle{empty}

\begin{center}

{\large\bfseries Can Generative AI help people navigate Radical Moral\\
Disagreements? The CONSIDER prototype}

\vspace{1.5em}

William Hohnen-Ford\textsuperscript{1*}, Sarah Chen\textsuperscript{1\dag}, Kathryn B.\ Francis\textsuperscript{1,2\dag}, Madeline G.\ Reinecke\textsuperscript{1,2\dag},\\
Ilina Singh\textsuperscript{1}, David Lyreskog\textsuperscript{1}.

\vspace{2em}

{\large Affiliations}

\vspace{0.8em}

\textsuperscript{1}University of Oxford, Department of Psychiatry

\vspace{0.4em}

\textsuperscript{2}University of Oxford, Uehiro Oxford Institute

\vspace{1.5em}

*Correspondence: \href{mailto:billy.hohnen-ford@psych.ox.ac.uk}{billy.hohnen-ford@psych.ox.ac.uk}

\vspace{1em}

\textsuperscript{\dag} These authors contributed equally

\end{center}

\vspace{1.5em}

\noindent\textbf{Funding:} \textbf{WHF}, \textbf{KF}, \textbf{MR}, \textbf{IS} and \textbf{DL} are funded by a Wellcome Trust Discovery Research Platform award [226801/Z/22/Z; Singh, PI.] This research was also supported by the Cosmos Institute and The Foundation for Individual Rights and Expression (\textbf{DL}, PI). 

\vspace{1.5em}

\noindent\textbf{Competing Interests:} None declared.

\vfill

\clearpage
\thispagestyle{empty}

\vspace{1.2em}

\begin{center}
\textbf{\large Abstract}
\end{center}

\vspace{0.6em}

\noindent
Radical Moral Disagreements (RMDs) are highly polarising topics that are increasingly censored in everyday life, with growing evidence suggesting that this polarisation carries measurable costs to public mental health. To address these challenges, some researchers have proposed Large Language Models (LLMs) as a means to support more democratic deliberation and better moral reasoning. Yet existing tools are poorly calibrated to help people navigate RMDs, because of their intense and divisive characteristics. This paper introduces CONSIDER, a prototype for a one-to-one AI tool for RMD navigation. Drawing on Mill’s account of the epistemic value of disagreement, CONSIDER aims at value clarification through structured disagreement with an opposing LLM-generated opinion. We describe CONSIDER’s design logic and analyse potential risks posed by such tools to guide future development.

\vspace{1em}


\clearpage
\pagestyle{plain}

\begin{center}
    {\Large\bfseries Introduction}
\end{center}

There is concern that generative AI may disrupt democracy by amplifying misinformation, disrupting elections, and further polarising society.\cite{summerfield2025} Yet the powerful language-processing and generative capabilities of Large Language Models (LLMs) have also enabled researchers to build generative AI-powered tools and interventions that may support ethical deliberation in new ways.\cite{tessler2026can} These include systems designed to help people find common ground,\cite{tessler2024} make key democratic functions such as citizens’ assemblies more accessible and inclusive,\cite{mckinney2024} and facilitate more constructive disagreement.\cite{argyle2023}

The emergence of these interventions coincides with broader social and political trends, marked by rising affective polarisation, increasing social distance, and decreasing mutual understanding between political groups.\cite{mutz2025} As these divisions deepen, citizens have become more reluctant to engage in cross-ideological moral discussion.\cite{iyengar2019,goel2025} Existing research suggests that this reluctance stems largely from the rising social, reputational, and emotional costs associated with expressing one's views on divisive topics.\cite{boland2024} This decline in willingness to engage across moral and political differences presents a serious challenge for democratic society. The ability to make collective progress on foundational moral and political questions largely depends on citizens’ willingness to confront and engage with opposing viewpoints.\cite{willis2026} When that willingness erodes, it threatens the quality of public moral reasoning, as well as the democratic processes through which societies negotiate moral disagreement.\cite{gutmann1995}

This paper illustrates a practical approach to exploring how purpose-built LLM-based tools can help people navigate divisive moral issues. We make four main contributions to the literature on how people can “do ethics with AI”, fostering improved ethical deliberation\footnote{Throughout this paper, we use ‘deliberation’ in the broad sense of structured moral reasoning directed at reaching well-supported answers to moral questions through active engagement with relevant reasons.\cite{richardson2018}} in public ethics contexts. First, we examine existing approaches to AI-assisted deliberation, arguing that they are poorly calibrated for ‘Radical Moral Disagreements’ (RMD) — highly polarising issues that resist conventional dialogue and democratic processes.\cite{lyreskog2025} Second, we propose a philosophically-grounded approach for AI-assisted RMD engagement, drawing on Mill's account of the epistemic value of disagreement. Third, we instantiate this approach through CONSIDER, an LLM-powered prototype for one-to-one disagreement navigation. We describe its design logic and the architectural choices that follow from our approach, answering calls in the literature for digital tools which provide ‘low-risk epistemic spaces’ for disagreement navigation — environments that reduce the social and reputational costs of engaging with opposing views while preserving genuine epistemic challenge.\cite{willis2026} Fourth, we analyse a broad range of ethical, social and psychological risks that AI-assisted RMD navigation tools may introduce, arguing that responsible development in this space requires confronting risks directly (rather than treating them as implementation details to be resolved after deployment).  
\vspace{1em}

\begin{center}
    {\Large\bfseries Section 1: The problem of Radical Moral Disagreement}
\end{center}
\subsection*{1a: Radical Moral Disagreement }
Lyreskog et al. (2025) conceptualise ‘Radical Moral Disagreements’ (RMDs) as a distinct social and philosophical phenomenon, made up of seven defining characteristics that distinguish them from more tractable moral disputes (See Table 1).\cite{lyreskog2025} Whether a particular disagreement becomes radical depends substantially on the social and political context in which it is realised. This is to say that the same underlying value conflict can be navigated productively or destructively depending on the conversational mode in which it is engaged.

Two features of RMDs are especially important for the discussion that follows. First, we spotlight intensity: RMDs are often identity-relevant in a way that distinguishes them from ordinary policy disagreements, making genuine engagement with opposing views feel personally threatening. Social Identity Theory posits that an individual’s ‘self-concept’ is shaped largely by the groups that they belong to (e.g., political affiliations, cultural communities). Insofar as these identities are woven into one’s sense of self, individuals may be motivated to see identity-relevant groups positively compared to others.\cite{tajfel2001} As Hogg and Smith (2007) argue, attitudes towards social and political issues function as group norms that help define social identity within a group context. When a particular social identity becomes salient, individuals do not simply independently express personal opinions but internalise attitudes that reflect the perceived values of the ingroup.\cite{hogg2010} In the context of RMDs, where moral positions on divisive issues (e.g., abortion, immigration) often function as markers of group membership, challenges to one’s position may therefore be experienced as threats to core features of social identity and sense of self. Take recent empirical work by Strickler (2018), where they find that the social identity dimension of partisanship drives anti-deliberative attitudes toward out-group disagreement.\cite{strickler2018} This suggests that the challenge of productive engagement with RMDs may lie not only in the content of people’s beliefs, but also in identity-relevant social dynamics. When moral and political positions become tied to group identity, people may experience opposing perspectives as threatening, both to the group and to the self. 

\vspace{1em}
\begin{table}[H]
\centering
\begin{tabular}{@{} l p{0.65\textwidth} @{}}
\toprule
\textbf{RMD Characteristic} & \textbf{Description} \\
\midrule
Collision & \rule{0pt}{2.6ex}\textit{Incompatible moral views come into direct conflict, particularly where collective decisions or policies are required.} \\
Intransigence & \rule{0pt}{2.6ex}\textit{Disagreements persist even under conditions favourable to dialogue, deliberation, and resolution.} \\
Intensity & \rule{0pt}{2.6ex}\textit{Moral disagreements generate strong emotional, existential, and identity-based responses.} \\
Polarisation & \rule{0pt}{2.6ex}\textit{Positions increasingly gravitate toward binary and oppositional extremes over time.} \\
Censure & \rule{0pt}{2.6ex}\textit{Public discussion becomes constrained through censorship, de-platforming, stigma, or suppression of opposing views.} \\
Social Disruption & \rule{0pt}{2.6ex}\textit{Disagreements extend beyond individuals and disrupt communities, institutions, and wider social cohesion.} \\
Contested Epistemologies & \rule{0pt}{2.6ex}\textit{Parties disagree not only about values, but also about evidence, expertise, and what counts as trustworthy knowledge.} \\
\bottomrule
\end{tabular}
\caption*{\textit{Table 1: Core characteristics of Radical Moral Disagreements (RMDs), adapted from Lyreskog et al., 2025. \cite{lyreskog2025}}}
\label{tab:rmd}
\end{table}

Second, we highlight polarisation: RMDs typically involve cases where there is no principled middle ground at which opposing parties can be expected to arrive. This claim is both descriptive and normative. Descriptively, RMDs resist convergence via a ‘middle ground’, because these disagreements are defined by differences in the foundational values. Fogelin (1985) describes this as clashes in ‘framework propositions’, where two parties may lack the shared assumptions and epistemic standards needed to make epistemic progress on their disagreement.\cite{fogelin2005} Normatively, while convergence toward a middle ground may be a realistic and desirable outcome for some political disagreements, RMDs are partially defined by the fact that multiple coherent moral frameworks can support opposing positions on the same issue. Combined with the identity-relevance described above, this means that expecting convergence as the outcome of engagement is neither realistic nor, in many cases, appropriate. This raises the question of what an AI-powered deliberative tool should be optimised to achieve. If not convergence of opinions, then what? We return to this question in Sections 2 and 3. 

\subsection*{1b: The Costs and Benefits of RMD Engagement }
The features outlined above make engagement with RMDs socially costly. Several dynamics help explain why. Affective polarisation, defined as the hostility and distrust between opposing political groups, has risen substantially over recent decades, \cite{iyengar2019, gidron2023} and research suggests that this hostility is driven more by the social identity dimensions of political affiliation than by substantive policy disagreement.\cite{mason2018} This is compounded by the moralisation of political discourse, in which attitudes held with moral conviction evoke characteristically negative emotions toward those who disagree.\cite{ryan2014,westheuser2026} These dynamics converge most intensely on the issues that exhibit RMD characteristics (e.g., issues that are identity-relevant, morally charged, and resistant to compromise). The result of this is ‘socially fraught disagreement’: cases where productive engagement is possible in principle but avoided in practice because the anticipated social, emotional, and reputational costs of genuine engagement with these issues have become too high.\cite{willis2026}  

These compounding factors may drive people to withdraw from engaging with RMDs altogether. In fact, greater affective polarisation has been shown to be associated with rises in self-censorship,\cite{gibson2023} suggesting that social threat can lead people to withdraw from identity-relevant discussion rather than engage it.\cite{boland2024} Additionally, the psychological costs of these dynamics may create powerful incentives for withdrawal, with the polarisation and politicisation of moral disagreements associated with heightened risks of anxiety and depression.\cite{nayak2021,ford2023} Lyreskog et al. (2025) identify censure as a defining characteristic of RMDs, highlighting that as disagreement becomes more radical, it becomes increasingly difficult to discuss in the public sphere, with one or both sides seeking to limit, stigmatise, or deny platforms to those with opposing views.\cite{lyreskog2025} Crucially, censure does not operate in a single mode. Nguyen (2020) distinguishes between epistemic bubbles, where opposing views are absent entirely and beliefs go unexamined due to lack of challenge, and echo chambers, where opposing views are encountered but have been actively delegitimised.\cite{nguyen2020} In the case of the latter, social media platforms may expose citizens to opposing positions but disproportionately amplify morally and emotionally charged content within ideological networks,\cite{brady2017} framing these positions as objects of ridicule or outrage rather than as warranting serious consideration. Both epistemic bubbles and echo chambers alike may result in similar outcomes — that individuals may be disinclined to reflect on the justification for their RMD-related viewpoints (or viewpoints that do not align with their own). This is to say that their beliefs may become dead dogma,\cite{mill1998} where a belief is endorsed out of habit and defended automatically, rather than the product of deliberation and careful thought. 

Importantly, however, RMDs are not inherently destructive and may catalyse moral progress in certain circumstances. In fact, some argue that engaging with divisive moral topics is a necessary condition for moral progress, forcing societies to confront the shallowness of existing understanding rather than leaving views unexamined.\cite{lane2024,moodyadams1999} Others argue that engagement with divisive issues is central to a functioning democracy, and that collective decisions hold legitimacy insofar as they are the outcome of deliberation among citizens who engage with opposing perspectives.\cite{gutmann1995} A paradox of sorts follows from this: RMDs are especially difficult to engage with in modern society, and yet hold deep democratic and moral value. Therefore, the challenge is not to eliminate engagement with RMDs, but rather to create conditions under which they can be engaged with productively and without psychological harm. Researchers have identified this tension as an urgent priority for further study, calling for the development and evaluation of new interventions tailored to RMDs, including digital tools designed to support rather than undermine constructive engagement with RMDs.\cite{lyreskog2025} It is in direct response to this challenge that we explore whether purpose-built LLM-based tools can help individuals navigate RMDs, using AI to support the process of individual ethical deliberation on contested moral topics.  
\vspace{1em}

\begin{center}
    {\Large\bfseries Section 2: Existing Approaches and Their Limits}
\end{center}
\subsection*{2a: Current AI Deliberation Tools  }
The capability of LLMs to process and generate large quantities of text, as well as their ability to be trained for a range of specialised usages, introduces a range of possible deliberative applications. These include content summarisation,\cite{small2023opportunities} consensus building,\cite{tessler2024} democratic mediation,\cite{li2025moderation} and moral guidance.\cite{giubilini2024} Such approaches to AI-assisted democratic deliberation can be broadly categorised into three types, distinguished by their interaction structure. First, there are multi-user platforms, which bring opposing groups into AI-mediated dialogue, deploying LLMs as moderation tools that rephrase emotionally charged statements,\cite{argyle2023} inject missing perspectives,\cite{fulay2025empty} or synthesise collective positions.\cite{small2023opportunities} Second, there are hybrid systems, which combine multi-user interaction with AI-driven synthesis. For instance, one prominent example is the Habermas Machine, which uses LLMs to iteratively generate and refine group statements that aim to represent the collective view of participants while preserving minority critiques.\cite{tessler2024} Finally, there are one-to-one LLM systems, which are specialised LLM-based chatbots designed for individual users to engage with directly.\cite{ma2025} We focus here on one-to-one LLM systems, chiefly because this is the primary mode of AI engagement in contemporary society.

We now turn to two existing approaches within one-to-one LLM systems: depolarisation tools and artificial moral advisor tools. We argue that while both are philosophically motivated and offer useful insights, neither is adequately calibrated for helping people to address the core challenge that RMDs pose: creating conditions under which genuine engagement with moral difference is possible, despite the social and identity-based costs that currently discourage it.  

The depolarisation approach takes political polarisation as its primary target, treating AI as an instrument for reducing attitudinal extremity. For instance, several studies use AI chatbots to move participants toward the midpoint of Likert scales on polarising political topics, operationalising success as reduced positional extremity.\cite{walter2025,hruschka2026} However, these approaches raise the question of whether moderation constitutes a meaningful goal. Here, a distinction can be made between issue polarisation, defined as the extremity of positional attitudes, and affective polarisation, defined as the hostility toward and dehumanisation of those who hold opposing views.\cite{iyengar2019} Both can carry democratic costs: extreme positional divergence can obstruct collective decision-making and compromise legislation when the distribution of views leaves little common ground, and affective polarisation can reduce cross-partisan cooperation and can create a culture of viewing political opponents as enemies rather than fellow citizens with differing values (ibid). Therefore, a tool that reduces positional extremity only targets part of the problem. Moreover, a society filled with more moderate viewpoints does not necessarily imply a better environment for productive and genuine ethical thinking; hostility can still exist in important debates, even if the differences in viewpoints are more subtle.\cite{schedler2023} Finally, there is also an ambiguity in what positional scale movement actually captures. A participant who moves from 7 to 4 on an immigration scale after a brief AI interaction may have genuinely reconsidered their position, or may have been subject to social influence, conversational fatigue, or deference to an authoritative interlocutor. The measure cannot distinguish between these, and optimising for it risks treating weakened conviction as epistemic improvement.  

An alternative approach conceptualises AI-assisted deliberation tools as 'Artificial Moral Advisors' (AMAs), designed to help users reason more carefully within their own moral frameworks. This approach argues that humans routinely fall short of their own moral standards not because their values are wrong but because of cognitive limitations, emotional biases, and the complexity of moral decisions, and that AI can help users reason more carefully from their own stated principles and improve moral decision-making.\cite{giubilini2018} Landes and colleagues (2025) distinguish between AI systems that act as moral authorities and those that function as 'inspiration machines', developing the user's own capacity for moral reasoning through Socratic questioning.\cite{landes2025} Leuenberger (2025) raises a complementary concern, arguing that personalised AI advisors risk narrowing the space for moral development by extrapolating conservatively from a user's existing commitments, and that a generalist advisor exposing users to competing viewpoints is preferable.\cite{leuenberger2025} Both converge on the principle that AI should improve the conditions for moral reasoning rather than resolve its outcomes, reinforced by the notion that genuine value conflicts are real features of the normative domain rather than errors to be corrected, and that AI can at best illuminate the structure of such conflicts while leaving their resolution to the agent.\cite{queloz2025} 

However, when applied to RMDs, this approach becomes insufficient. Fundamentally, AMA tools address a problem internal to the individual reasoner — poor reasoning, inconsistency, failure to act on one's own values. Yet for RMDs, these are genuine conflicts between coherent but incompatible value frameworks. Our goal is therefore not only to improve a user’s deliberative coherence or help them reflect more carefully on their own commitments (a framework for which the AMA literature provides) but to give them an epistemically low-risk space to engage with a coherent opposing view that contests those commitments on their own terms, addressing the socially fraught nature of these disagreements that is making them increasingly unavailable in contemporary discourse.\cite{willis2026}

\subsection*{2b: The Goal Problem  }

The approaches reviewed above are both, broadly, AI-assisted deliberation tools, but they are aligned to different proxies for deliberative success. Depolarisation tools target positional extremity and see a successful outcome as moderating people’s viewpoints on a Likert scale. Common-ground systems such as the Habermas Machine avoid this midpoint assumption, aligned instead to the production of collectively endorsed group statements, with explicit attention to minority critiques and fair representation.\cite{tessler2024} AMA systems are different again, targeting the quality of individual moral reasoning. Each of these targets is defensible in certain contexts, but none addresses the key deliberative challenges RMDs raise. This can be framed as an iteration of the AI alignment problem, referring to the challenge of specifying human values with sufficient precision that AI systems optimise for what we actually care about, rather than for proxy measures that correlate imperfectly with our goals.\cite{gabriel2020} This raises the question: What should a tool like this be trying to achieve, and why is that the proper goal? We call this ‘the goal problem’. In what comes, we demonstrate how CONSIDER speaks to the goal problem, as posed by RMDs.
\vspace{1em}

\begin{center}
    {\Large\bfseries Section 3: A Millian Framework for AI-Assisted Engagement with RMDs}
\end{center}

In this section, we address ‘the goal problem’ in our AI-assisted deliberative tool and draw on Mill’s account of the epistemic value of disagreement to argue that AI-assisted engagement with RMDs should aim at value clarification through structured disagreement. 
In \textit{On Liberty}, Mill offers a threefold justification for engaging seriously with opposing views, regardless of whether those views are correct. It describes three possible epistemic outcomes of genuine engagement with disagreement:  

\begin{enumerate}
    \item The opposing view may be correct, and suppressing or avoiding it risks forfeiting access to the truth.
    \item Even if the opposing view is wrong, it may contain a partial truth that the dominant view lacks.
    \item Even if the opposing view is wholly wrong, genuine engagement with it clarifies and deepens one's understanding of one's own position.
\end{enumerate}

Without challenge through one of these possibilities, Mill argues, beliefs become dead dogma rather than living conviction,\cite{mill1998} held by habit and reflex rather than genuinely understood. 

We adopt Mill as a design framework specifically calibrated to the features of RMDs established in Section 1. All three epistemic benefits obtain without requiring either party to converge on a shared position, making the framework appropriate for disagreements where there is frequently no principled middle ground. Mill's warning against dead dogma directly describes the withdrawal dynamics identified in Section 1b, in which disengagement from RMDs leaves moral commitments unexamined. Unlike the approaches reviewed in Section 2, a Millian tool does not treat moderation as success or limit itself to improving individual reasoning; its core function is to provide structured engagement with a coherent opposing moral framework, giving users a low-risk space to do something that holds both moral and democratic value but has become increasingly difficult in practice.\cite{willis2026}

Crucially, such a tool should address both the problem of epistemic bubbles and echo chambers.\cite{nguyen2020} A one-to-one LLM tool responds to the first problem by making opposing views available via disagreement. Yet the problem of echo chambers is more challenging, where opposing viewpoints are dismissed. To address this, a Millian tool must present opposing views in a way that is serious enough to challenge the user, yet not so antagonistic that it reproduces the dynamics that make RMD engagement socially and psychologically costly in the first place.

\vspace{1em}

\begin{center}
    {\Large\bfseries Section 4: The CONSIDER tool and its rationale }
\end{center}

We now turn to CONSIDER, a web-based prototype designed to help individuals engage with RMDs through a structured, multi-stage interaction grounded in the Millian approach described above. It is implemented using a multi-agent architecture in which separate AI agents (Llama-3) handle distinct stages of the interaction. The full prompt instructions and codebase are publicly available on Github  \footnote{https://github.com/BillyhfPsyc/CONSIDER}. We anticipate making the system available under an open-source licence following further development and evaluation; in the meantime, researchers wishing to adapt or build upon the system are welcome to contact the corresponding author. Here, we outline the flow of the CONSIDER prototype (See Figure 1), with design choices and rationale.

\subsection*{\textit {Stage One: Topic Selection}}
The interaction begins with a topic selection from a curated set of RMD-relevant issues (such as immigration, abortion, or war). Rather than inviting broad conceptual discussion, the AI invites the user to specify a concrete aspect of the topic they wish to focus on, such as the moral justifiability of a particular action within a specific military conflict. This design choice reflects the concern that abstract framing may reduce the felt reality of a disagreement and decrease the likelihood of genuine engagement.\cite{pavarini2021}

\begin{figure}[H]
    \centering
    \includegraphics[width=\textwidth]{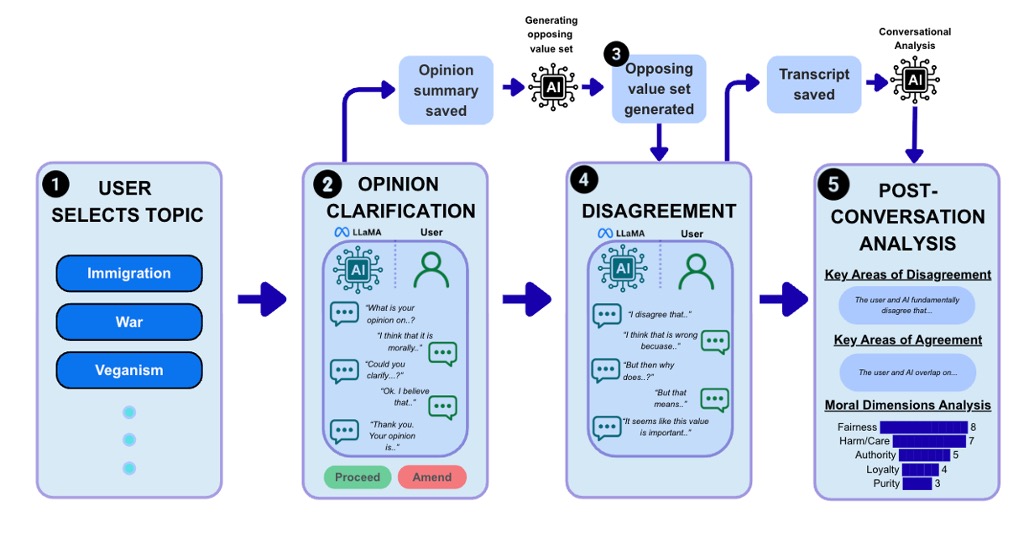}
    \textit{\caption{\textit{Overview of the CONSIDER prototype interaction flow. The system guides users through five stages: (1) topic selection from a curated set of RMD-relevant issues; (2) opinion clarification, in which an AI agent elicits and summarises the user's position through dialogue; (3) opposing value set generation, in which a separate AI agent produces a stable opposing profile based on the user's confirmed summary; (4) structured disagreement, in which a disagreement agent engages the user in sustained conversation from the opposing perspective; and (5) post-conversation analysis, in which the transcript is analysed across multiple dimensions.}}}
    \label{fig:consider}
\end{figure}

\subsection*{\textit {Stage Two: Opinion Calibration}  }
Following topic selection, the user engages in a one-to-one discussion with an AI agent, instructed to gather three elements from the user: (1) their core belief on the topic stated in their own words, (2) the values or reasons underlying that belief, and (3) how strongly they hold this belief. If aspects of the user’s beliefs are vague or unclear, the agent asks clarifying questions to specify the user’s position. Once sufficient information has been gathered, the agent produces a structured summary of the user's position, which is presented back to the user for confirmation or amendment. This serves a multitude of functions. As outlined in the AMA literature, articulating one's own commitments with precision is itself a form of ethical deliberation.\cite{landes2025} Additionally, it also contributes toward the goal of value clarification. A user who has not first reflected on what they believe and why is poorly positioned to learn anything from encountering a challenge to those beliefs. The confirmed summary is saved and passed forward to subsequent stages as a stable representation of the user's view. 

\subsection*{\textit {Stage Three: Opposing Profile Generation}}
Based on this summarisation, a separate AI agent generates an ‘opposition value profile’ which acts as a stable representation of a realistic position that contests that of the user. The profile is constructed around core values, non-negotiable commitments, specific claims the opposing value profile would advance, and the principal reason for rejecting the user's stance. Notably, this separation of tasks across different agents reflects a broader design principle in multi-agent LLM systems. Complex interactions can be made more reliable by dividing them into narrower stages, where each agent performs a specific task and passes a structured output to the next stage. \cite{hong2024,zhou2025approach} In CONSIDER, the structured output is the opposing value profile, with the purpose of giving the subsequent disagreement agent a clear and stable position to work from, rather than allowing the opposing view to be gradually shaped by the user’s wording, assumptions, or framing during the live conversation. Without this separation, there is a risk that the opposing position would be shaped by sycophantic tendencies or by mirroring the user's language, producing a version of disagreement that is a reactive inversion of the user’s framing instead of a consistent position.  

\subsection*{\textit {Stage Four: Structured Disagreement}}
The confirmed user summary and the generated opposing profile are passed to a disagreement agent, which conducts a sustained conversation with the user from the perspective of the opposing value set. Several conversational techniques are built into the agent's instructions to support the goal of value clarification rather than winning a debate. The agent gives reasons grounded in the opposing value set when presenting counterarguments, rather than producing generic objections; it reflects the user's own reasoning back to them and asks clarifying questions where the user's position is ambiguous; and it works to pinpoint where the disagreement between the two positions lies, such as distinguishing empirical disputes (disagreements about whether X causes Y) from foundational value conflicts (disagreements about which moral commitments should take priority).
Drawing on the AMA literature's insight that AI should improve the conditions for reasoning rather than resolve its outcomes,\cite{giubilini2018} the disagreement agent also integrates reflective questioning and clarification of the user's own reasoning alongside the structured challenge. This addresses the tension between presenting opposing views seriously enough to challenge the user, and tempering that challenge so it does not reproduce the antagonistic dynamics that make RMD engagement socially costly in the first place. 

\subsection*{\textit {Stage Five: Analysis and Reflection}}
Following the structured disagreement, the conversation transcript is passed to a separate analysis LLM that generates a reflective summary for the user. This stage is designed to support the value clarification goal by helping users step back from the live interaction and examine what the disagreement revealed. Multiple LLM instances are used to analyse the transcript across several dimensions. First, areas of agreement, disagreement, and potential agreement (positions that were not discussed but where common ground may exist) are extracted and presented to the user, encouraging reflection on the overall structure of the exchange. Second, a separate model assesses the user’s ‘epistemic humility’, analysing how far they acknowledged uncertainty, recognised the limits of their knowledge, and engaged seriously with counterarguments. Third, an LLM-as-judge approach is used to evaluate how strongly particular moral dimensions (such as autonomy, harm, or fairness) featured in the discussion, with supporting evidence drawn from the transcript and reasons given for each assessment.\cite{rathje2024} These results are presented to the user in a visual format designed to be accessible and engaging.  

The inclusion of this reflective stage serves two purposes. First, it provides users with a structured account of what happened in the conversation that they can return to. Second, it provides an analysis of the disagreement designed to encourage reflection, highlighting where positions converged, where they remained apart, and what kinds of moral reasoning were at stake. This post-conversation reflection is intended to consolidate whatever value clarification occurred during the live interaction, and to prompt further thinking that the user can carry into subsequent conversations, whether with AI or with other people.

\vspace{1em}

\begin{center}
    {\Large\bfseries Section 5: Risks and Prototype Limitations}
\end{center}

The preceding sections have outlined the problem of RMDs, critically examined existing approaches to AI-assisted deliberation, proposed a Millian approach for AI-assisted RMD engagement, and instantiated this approach through the CONSIDER prototype. In this final section, we turn to the various risks that tools of this kind may pose (See Table 2), the limitations of our current work, and directions for future research. The risks discussed below emerged from an interdisciplinary expert seminar held with sixteen experts who tested CONSIDER in February 2026.\cite{hohnenford2026} While each of the points raised below warrants extended analysis, our aim here is to map the terrain of ethical and practical challenges that developers and researchers in this space must confront. The risks discussed below operate across ethical, social and psychological dimensions, and many cut across these categories in practice. We therefore discuss them together rather than imposing a strict taxonomy.

\subsection*{5a: Risks}
The first cluster of risks concerns human autonomy. ‘Displacement’ refers to the concern that CONSIDER will not supplement real-world RMD engagement but progressively replace it. The features that make it epistemically low-risk - such as the absence of social costs, the consistency of the AI interlocutor, and the user's control over the interaction - may simultaneously render it preferable to the harder work of human-to-human conversation. Empirical work on human–chatbot interaction suggests that users experience less fear of judgement when interacting with chatbots than with human interlocutors,\cite{croes2024} while emerging work on companion AI warns that such systems may replace or deskill human relationships unless deliberately designed as bridges back to human interaction.\cite{malfacini2025} CONSIDER is therefore framed as a preparatory tool intended to build deliberative capacity for real-world engagement. 

A related concern is belief offloading, which theorises that if users repeatedly engage with an AI that articulates their values back to them, the site of moral reasoning may shift from the individual to the machine, attenuating the deliberative capacity the tool is intended to strengthen.\cite{guingrich2026belief} This is supported by recent empirical work which demonstrated that such repeated interaction may shape users' values through the system's framing and vocabulary in ways that are not detected by the user.\cite{li2026} In CONSIDER, this risk is most concentrated in the opinion calibration stage, where the AI produces a structured summary of the user's position. Over repeated use, users might come to adopt the system's framing and vocabulary rather than generating their own. CONSIDER’s confirmation step provides a partial check on this issue, requiring users to actively evaluate the AI’s articulation against their own understanding. However, the extent of belief offloading in everyday chatbot use remains an open empirical question.

This risk is compounded by concerns regarding the diversity of opposing perspectives that modern AI chatbots currently generate. Given documented political and moral biases in LLMs,\cite{rettenberger2025,abdulhai2024} there is a risk that the counter-perspectives produced reflect a narrow range of moral frameworks rather than the genuine plurality that exists on any given RMD. A variety of methodologies could address this risk, including the usage of political and moral bias evaluation methods\cite{bang2024} and the diversification of post-training populations.\cite{kirk2024} 

A second cluster of risks concerns the effects on users’ patterns of reasoning and engagement. For instance, one possible risk is that exposure to disagreement may, under some conditions, deepen convictions rather than open them to reflection, particularly when challenge is experienced purely as an attack on identity-relevant beliefs (what we term the "entrenchment paradox"). Research on motivated reasoning and identity-protective cognition demonstrates that counter-attitudinal arguments on identity-relevant topics can sometimes reinforce prior positions rather than dislodge them.\cite{kahan2013} Whether disagreements mediated through CONSIDER produce entrenchment, reflection, or greater tolerance for opposing views remains an open empirical question that future research should address.

A related risk is what we term ‘play-to-win engagement’, where users treat the interaction as a debate to be won rather than a disagreement to be understood. An AI that refuses to concede points may inadvertently encourage competitive rather than reflective engagement, while a sycophantic AI that concedes too readily fails to deliver the sustained challenge the approach requires. Calibrating appropriate assertiveness is an ongoing design challenge for the field. For instance, if CONSIDER's disagreement dynamics inadvertently reinforce zero-sum perceptions, this could further entrench the avoidance behaviour that the tool seeks to counteract.\cite{boland2024} CONSIDER tries to mitigate this possibility by pulling the user towards understanding, taking inspiration from AMA approaches, rather than disputing evidence in a purely adversarial manner (which may encourage a play-to-win approach).

\vspace{1em}
\begin{table}[H]
\centering
\begin{tabular}{@{} l p{0.65\textwidth} @{}}
\toprule
\textbf{RISK} & \textbf{Description} \\
\midrule
Displacement & \rule{0pt}{2.6ex}\textit{Users may come to prefer AI-mediated disagreement over real-world engagement because it is less socially and emotionally costly.} \\
Belief offloading & \rule{0pt}{2.6ex}\textit{Users may use the system to form, maintain, or revise moral beliefs, such that the AI becomes part of the mechanism by which those beliefs are adopted and sustained.} \\
Homogeneity of values & \rule{0pt}{2.6ex}\textit{The system may reproduce a narrow or biased range of moral perspectives, giving the appearance of pluralism without genuine diversity.} \\
Techno-solutionism & \rule{0pt}{2.6ex}\textit{The tool may encourage the view that deeply social and political problems can be addressed primarily through technical intervention.} \\
Entrenchment paradox & \rule{0pt}{2.6ex}\textit{Exposure to opposing arguments may strengthen users’ prior convictions rather than open them to reflection or reconsideration.} \\
Play-to-win engagement & \rule{0pt}{2.6ex}\textit{Users may approach the interaction as a contest to be won rather than a disagreement to be understood.} \\
 Psychological harm & \rule{0pt}{2.6ex}\textit{Identity-relevant disagreement may cause distress or exacerbate existing psychological vulnerabilities in some users.} \\
\bottomrule
\end{tabular}
\caption*{\textit{Table 2: Key risks associated with AI-assisted navigation of Radical Moral 	Disagreements. Adapted from our Expert Seminar on using AI for navigating Radical 	Moral Disagreement. \cite{hohnenford2026}}}
\label{tab:rmd}
\end{table}

Finally, we turn to safety risks, considering the possibility of harm to vulnerable users. While CONSIDER seeks to be a low-risk space for RMD navigation, certain model responses or behaviour may cause distress, given the sensitivity and identity-relevance of these topics. This is particularly the case for individuals with pre-existing mental health vulnerabilities, where such an experience may exacerbate existing difficulties rather than building deliberative capacity. The risk is compounded by people’s tendency to anthropomorphise AI systems,\cite{colombatto2024,reinecke2025} whereby users may experience pushback from the model as a personal affront. Moreover, the stochastic character of LLM outputs mean that even carefully prompted or fine-tuned systems may produce individual responses that are poorly calibrated to a user's emotional state. The extent and nature of such harm is an empirical question that is substantially determined by aspects of model behaviour that remain difficult to predict and control. Future development should incorporate safeguarding mechanisms, including distress detection and dynamic adjustment of interaction intensity. More broadly, the psychological effects of sustained AI-mediated moral disagreement remain poorly understood and require systematic investigation before tools of this kind are deployed at scale.

\subsection*{5b: Limitations}
The present research provides a preliminary demonstration, through the CONSIDER prototype, of how LLMs may help people navigate radical moral disagreements. As an early-stage prototype, however, this tool is subject to several important limitations. From a technical perspective, the prototype is built entirely through prompt engineering, relying on carefully constructed system prompts to shape model behaviour across each stage of the interaction. While prompt engineering can produce meaningful changes in model output,\cite{marvin2024} it provides limited control over deeper aspects of model behaviour.\cite{wang2025} Alternative methods include fine-tuning and neural vector steering,\cite{turner2023steering} potentially offering more reliable control over these behavioural tendencies.  Additional technical limitations include the absence of retrieval-augmented generation (RAG),\cite{lewis2020} meaning the chatbot cannot search for evidence online to support its claims with evidence outside their training data. This carries both epistemic and deliberative consequences: epistemically, the system cannot verify empirical claims or respond to emerging evidence, limiting its ability to engage with the factual dimensions of RMDs; deliberatively, without retrieval capabilities it risks recycling generic arguments or failing to challenge misinformation when users introduce it. This was partially addressed by instructing the model to engage with empirical evidence for instance without necessarily disputing the evidence (e.g., by being curious about it and probing it conditionally).  

Additionally, the prototype is designed around deliberative norms situated within the liberal tradition of individual autonomy and open debate. In contexts where the social meaning of disagreement, the distribution of moral authority, or the relationship between individual and collective reasoning differ substantially, a tool built on these assumptions may be counterproductive. Addressing this limitation requires co-design with diverse communities and sustained attention to the cultural presuppositions embedded in the tool's architecture. 

\vspace{1em}

\begin{center}
    {\Large\bfseries Conclusion and Future Directions}
\end{center}

This paper has examined the possibility of using purpose-built LLM-based tools to help individuals navigate RMDs. We have argued that existing approaches to AI-assisted deliberation, such as depolarisation tools and artificial moral advisors, are each aligned to normative targets that are poorly calibrated for RMDs, framing this as an instance of the broader AI alignment problem applied to the deliberation domain. Drawing on Mill's account of the epistemic value of disagreement, we proposed that such tools should aim at value clarification through structured disagreement: helping users understand the structure of their moral commitments and the points at which those commitments diverge from coherent opposing frameworks, without presupposing that convergence is either achievable or desirable. We instantiated this approach through CONSIDER, a prototype that translates these philosophical commitments into a set of architectural choices for a working system, while also introducing core risks in the development of such tools.  

There are several directions for future research to follow. Most immediately, empirical evaluation is needed to test whether CONSIDER produces the outcomes the Millian approach predicts: genuine value clarification, improved understanding of the structure of disagreement, and the development of deliberative capacities that transfer to real-world contexts. Such work should include both controlled experiments examining the tool's effects on users' understanding of their own and opposing positions, and longer-term studies tracking whether engagement with the tool affects users' willingness and ability to engage with RMDs in interpersonal settings. Beyond evaluation, the question of how tools like CONSIDER might be integrated into broader deliberative contexts deserves attention. One possibility is integrating RMD-navigation capabilities directly into general-purpose AI assistants, such that when a user engages with a morally contested topic in ordinary conversation, the system shifts into a more structured deliberative mode rather than defaulting to the sycophantic responses that current systems often produce.\cite{sharma2024} This would allow the deliberative benefits of structured disagreement to reach users who would not seek out a dedicated tool but who regularly encounter RMDs in their interactions with AI. 

CONSIDER remains a prototype and raises many empirical questions for future work to address. What the prototype demonstrates is that a theory-driven approach to AI-assisted deliberation on contested moral questions is both technically feasible and philosophically distinctive, and that the design space for such tools is richer than existing approaches have explored. Whether AI can reliably help people navigate RMDs is ultimately an empirical question. This paper has sought to clarify what such navigation could involve, and to identify the conditions under which it might succeed. 

\vspace{1em}

\textbf{Acknowledgements:} For the purpose of Open Access, the author has applied a CC BY public copyright licence to any Author Accepted Manuscript version arising from this submission. SC acknowledges the support of the Rhodes Trust. Additionally, we would like to express our gratitude to Dr. Adam Etinson for helpful discussions on Nguyen’s philosophy, Dr. Lucienne Spencer for helpful contributions, Georgina Kenny for assistance with the development of the figures, and all of the experts who contributed to our online seminar. 

\bibliographystyle{vancouver}
\bibliography{refs}          
\end{document}